\documentclass[final,1p,times]{elsarticle} 
\usepackage{graphicx} 
\usepackage{amssymb} 
\usepackage{amsthm} 
\usepackage{lineno} 

\usepackage{subfigure}
\journal{Nuclear Physics A} 

\begin{document} 

\begin{frontmatter} 


\title{In-Medium Modifications of Low-Mass Vector Mesons in PHENIX at RHIC }

\author{Yuji Tsuchimoto$^{a}$ for the PHENIX collaboration}

\address[a]{Graduate School of Science, Hiroshima University, 
1-3-1 Kagamiyama, Higashi-hiroshima, 739-8526, Japan}

\begin{abstract} 
Measurements at RHIC have established
the creation of a Quark Gluon Plasma (QGP)
in most central heavy-ion collisions.
An important tool to understand properties of the QGP
is study of the spectral shapes of low-mass vector mesons (LVM's),
$\rho$, $\omega$ and $\phi$,
which can be modified in the medium
by partial restoration of chiral symmetry.
This modification may be accessed directly
by measuring low-momentum LVM's via their decays into lepton pairs
inside the hot matter.
Since leptons are not subject to the strong interaction,
they do not rescatter on their way out of the medium.
The PHENIX experiment at RHIC has measured LVM production at mid-rapidity
in $p$+$p$, $d$+Au and Au+Au collisions at $\sqrt{s_{\rm NN}}$ = 200 GeV\@.
Mass peaks for the LVM's have been observed
in the di-electron invariant mass spectra
with a resolution of 10 MeV/$c^2$ in all of the three collision systems.
The extracted spectra, mass and width of $\omega$ and $\phi$
in $p$+$p$, $d$+Au and Au+Au,
in the leptonic and hadronic decay channels are reviewed.
As the widths of the mesons may be affected in the medium,
the branching ratios of various decay modes may also be modified
from the values in vacuum.
The relative branching ratio is compared between
$\phi\rightarrow e^+e^-$ and $\phi\rightarrow K^+K^-$,
which may be sensitive to the mass modification
due to the small Q-value of $\phi\rightarrow K^+K^-$.
\end{abstract} 

\end{frontmatter} 



\section{Introduction}

QCD is a non-abelian quantum gauge theory
which discribes the strong interaction.
The QCD predicts that a Quark Gluon Plasma (QGP) is created
at hot and/or dense nuclear matter.
Recently, there are many experimental results
suggesting that the QGP has been created
in heavy ion collisions at RHIC\@.
For the next task,
we are on stage to study the properties of the QGP\@.

Within the framework of QCD,
most of the ordinary hadronic mass is expected to be obtained
via the spontaneous breaking of chiral symmetry~\cite{Nambu:1961tp}.
This effective mass depends on the vacuum energy density,
and it will vanish in hot and/or dense matter.
Some model calculations predict mass modification
of low-mass vector mesons (LVM's)
in medium~\cite{Brown:1991kk,Rapp:1999ej,Cheng:2006qk,Aoki:2006br}.
The mass may be affected and become lighter
and the width may become wider in the hot medium.

LVM's have both leptonic and hadronic decay channels.
The hadronic decay channels have an advantage of larger branching ratios.
However, hadrons from deep within the hot matter are scattered
by the strong force,
thereby loosing information of the original decay.
On the other hand, as leptons are not subject to the strong interaction,
leptonic probes can carry information of the vector mesons in medium.
The most straight-forward and direct method to detect the mass modification is
line-shape analysis on invariant mass spectra
of daughter leptons of the LVM's.
KEK E325 collaboration has reported an excess
on the lighter side of the LVM peaks
on $e^+e^-$ invariant mass spectra~\cite{Muto:2005za}.
However, this line-shape analysis needs high statistics,
which is difficult to collect at collider experiments.

Another interesting probe is the decay branch
of $\phi\rightarrow K^+K^-$ to $\phi\rightarrow e^+e^-$.
Due to the small Q-value of $\phi\rightarrow K^+K^-$/${\bar K}^0K^0$,
the decay branch of $\phi$ may be sensitive
to the in-medium mass modification.
Some models predict
that mass of $\phi$ can be modified to be lighter
than mass of a pair of kaons
in hot and/or dense medium~ \cite{Hatsuda:1991ez},
in which extreme case,
$\phi$ cannot decay into a kaon pair.

\section{Experiment}

The LVM measurements presented in this contribution were obtained
from the data samples accumulated by the PHENIX experiment
during $p$+$p$, $d$+Au and Au+Au collisions
at $\sqrt{s_{\rm NN}}$ = 200~GeV/$c$
in 2003-2005 physics runs.
LVM's are measured in both of electronic and hadronic channels
using the PHENIX central spectrometer,
which measures fully identified
hadrons, electrons and photons \cite{Adcox:2003zp}.
Charged particles are tracked by drift chambers and pad chambers
which provide momentum of the charged particles.
Kaons are identified by the time of flight information
measured by ToF counters (Full-PID method).
For higher $p_{\rm T}$ kaons,
the $\phi$ can be also reconstructed
without kaon identification (no-PID method).
Electrons and positrons are identified
by {\v C}erenkov emmition in RICH detectors,
and by the ratio of momentum to energy
measured by electromagnetic calorimeters.
$\rho/\omega/\phi\rightarrow e^+e^-$ are reconstracted by $e^+e^-$ pairs.
Photon are measured by the calorimeters,
$\pi^0\rightarrow 2\gamma$ are reconstracted from 2$\gamma$,
and $\omega\rightarrow\pi^0\gamma$ are from 3$\gamma$.

\section{Results}

\begin{figure}[tb]
 \centering
 \begin{minipage}{0.32\hsize}
  \subfigure[]{
\includegraphics[clip,width=\hsize]{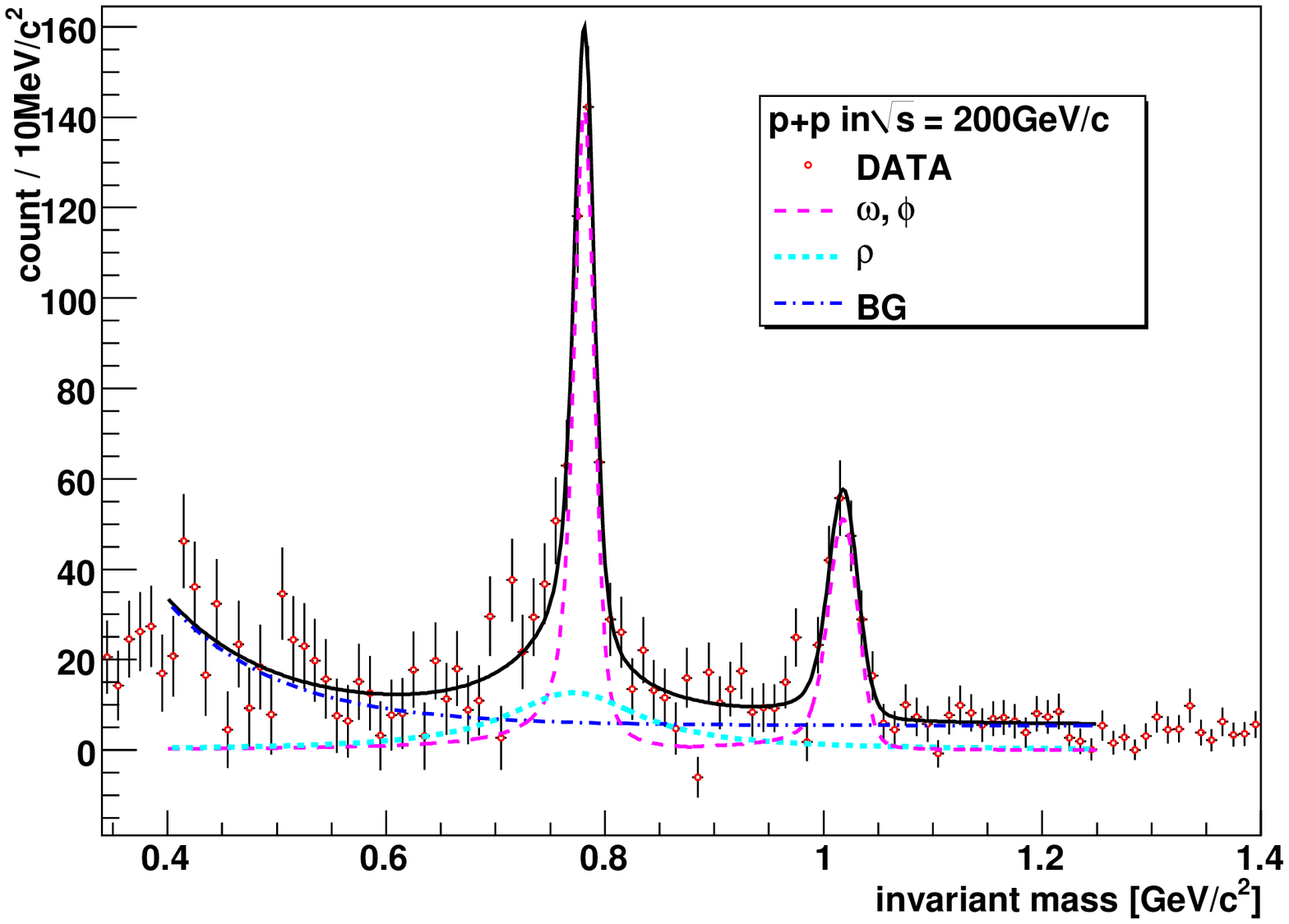}
  \label{fig:invmass-ee-pp-200-lowmass}
  }
 \end{minipage}
\begin{minipage}{0.32\hsize}
  \subfigure[]{
\includegraphics[clip,angle=270,clip,width=\hsize]{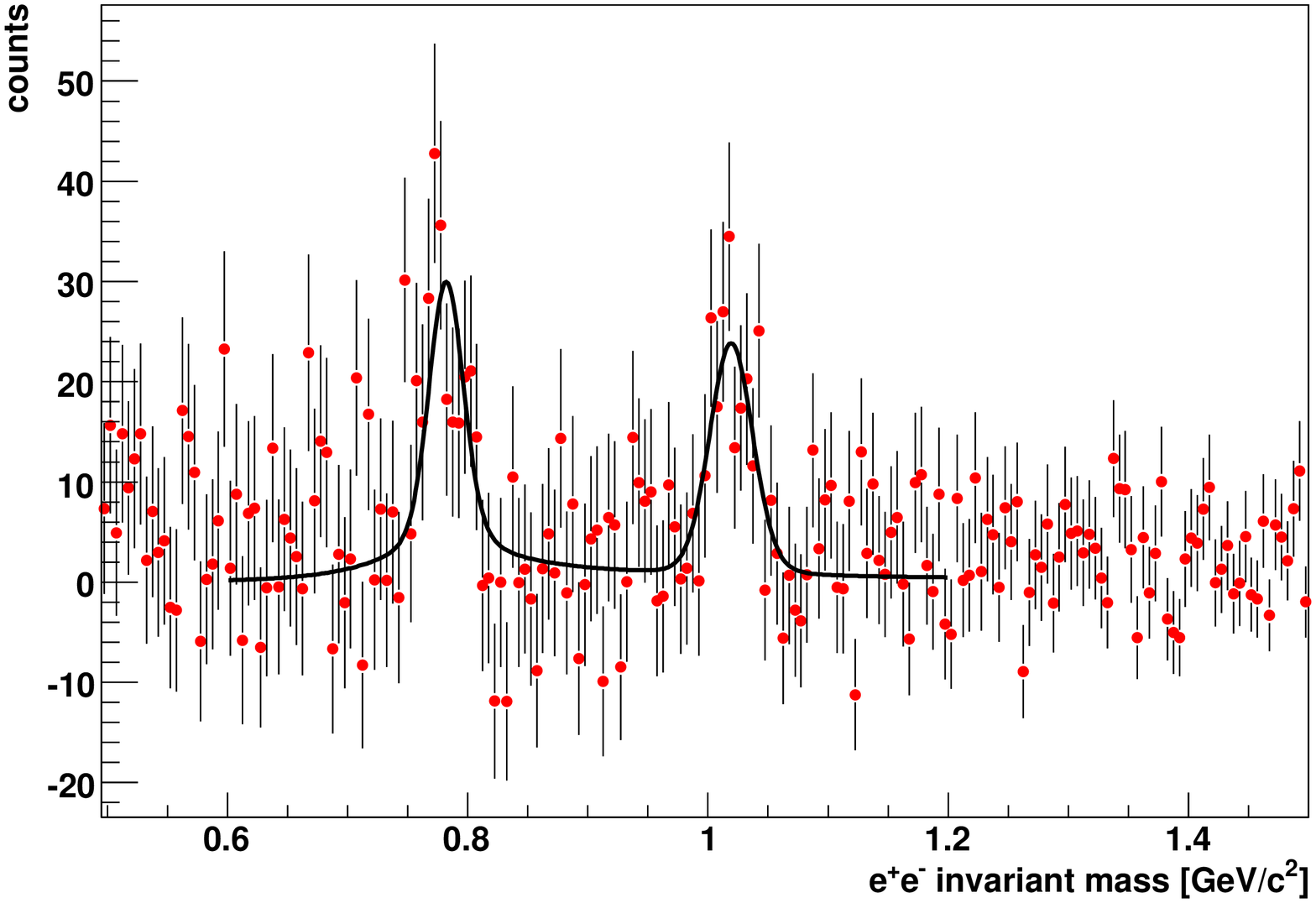}
  \label{fig:invmass-ee-dAu-200-lowmass}
 }
\end{minipage}
\begin{minipage}{0.32\hsize}
  \subfigure[]{
\includegraphics[clip,width=\linewidth]{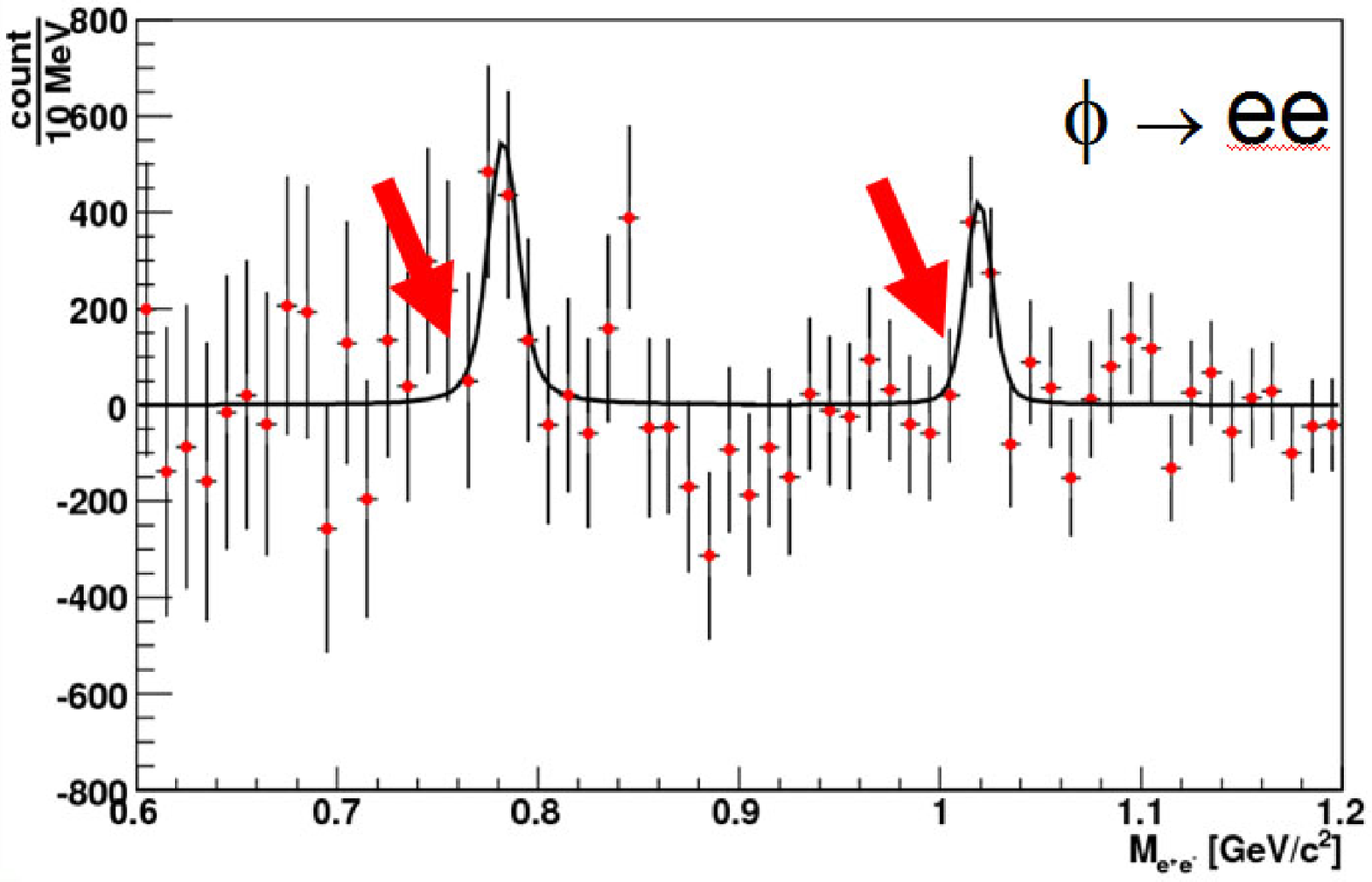}
  \label{fig:ee_AuAu_2}
}
\end{minipage}
\caption{
 Invariant mass of $e^+e^-$ pairs
 detected by the PHENIX central spectrometers
 in (a) $p$+$p$, (b) $d$+Au and (c) Au+Au collisions
 at $\sqrt{s_{\rm NN}}$ = 200~GeV.
 The $\omega$ and $\phi$ resonances clearly visible
 at 0.78 GeV/$c^2$ and 1.02 GeV/$c^2$ in these plots.
 Combinatorial backgrounds are subtracted by event mixing.  
The spectrum is fit to the $\phi$ and $\omega$ resonances where
the masses and widths are set to the PDG values;
 the Breit-Wigner resonance shape is corrected for the radiative tail; an additional Gaussian term is convoluted withe the Breit-Wigner width taken from simulations.
$\rho$ contribution is assumed as the the same yield as $\omega$, as shown as dotted line in (a).
 The residual continuum component is estimated by exponential fit as shown by dash-dotted line in (a).
}
\label{fig:invmass-ee-200-lowmass}
\end{figure}

\begin{figure}[tb]
 \centering
 \begin{minipage}{0.32\hsize}
  \subfigure[]{
\includegraphics[clip,width=\hsize,height=0.7\hsize]{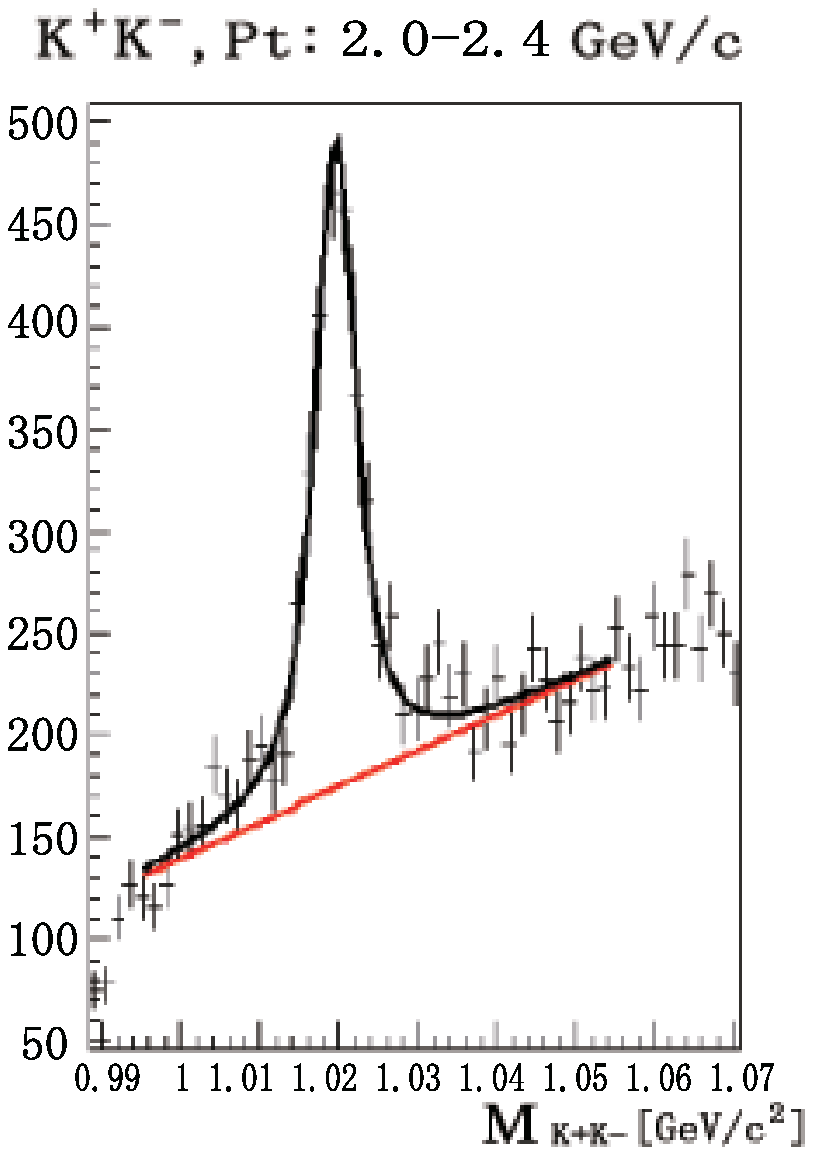}
  \label{fig:kk_pp}
  }
 \end{minipage}
 \begin{minipage}{0.32\hsize}
  \subfigure[]{
\includegraphics[clip,width=\hsize]{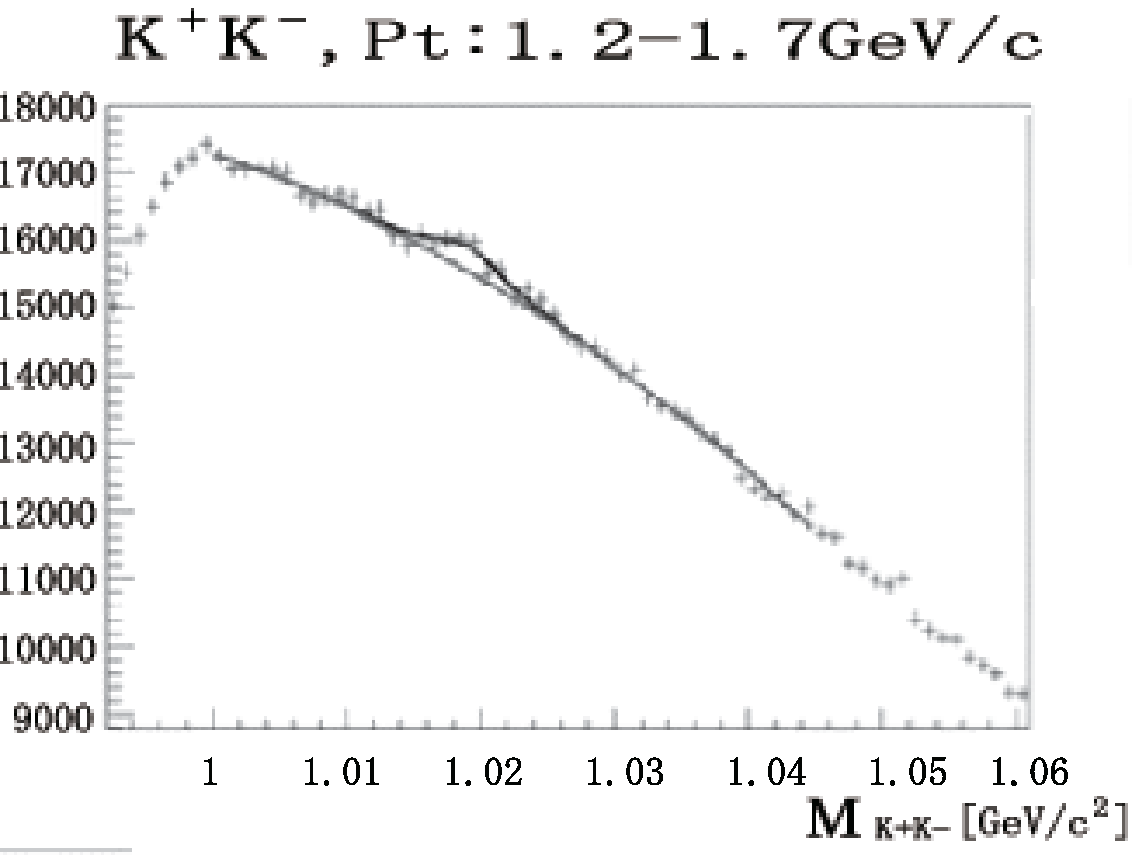}
  \label{fig:kk_dAu_1}
  }
 \end{minipage}
 \begin{minipage}{0.32\hsize}
  \subfigure[]{
\includegraphics[clip,width=\linewidth,height=0.8\hsize]{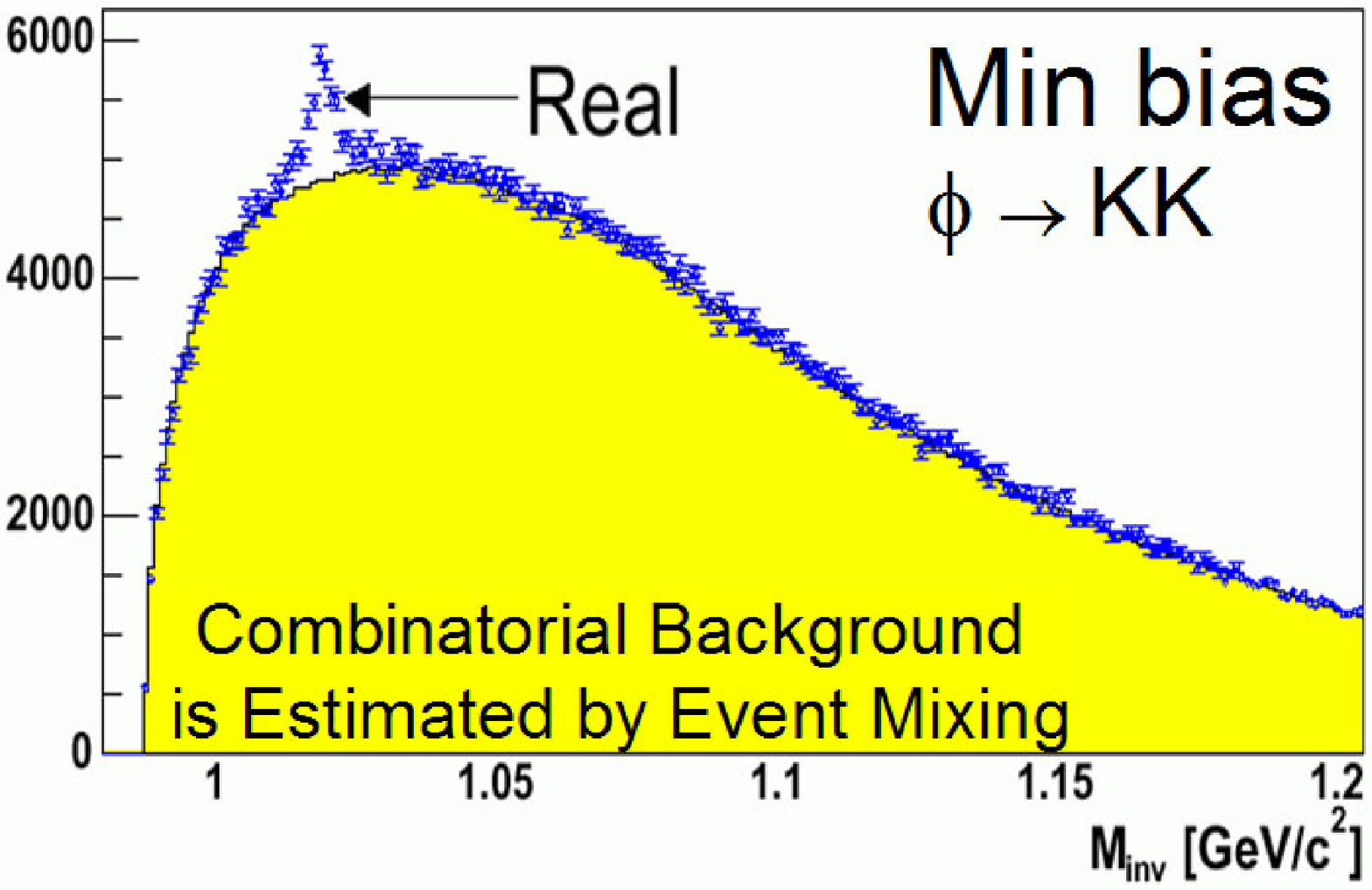}
  \label{fig:kk_AuAu}
  }
 \end{minipage}
\caption{
 Invariant mass of $K^+K^-$ pairs detected by the PHENIX central spectrometers in p+p (a), d+Au (b) and Au+Au (c) collisions at $\sqrt{s_{\rm NN}}$=200~GeV.
 The $\phi$ resonances clearly visible at 1.02~GeV/$c^2$ in these plots.
 Combinatorial backgrounds are estimated by event mixing.  
}
\label{fig:invmass-kk-200-lowmass}
\end{figure}

Figures \ref{fig:invmass-ee-200-lowmass} and \ref{fig:invmass-kk-200-lowmass}
show invariant mass spectra of $e^+e^-$ and $K^+K^-$
in (a) $p$+$p$, (b) $d$+Au and (c) Au+Au collisions, respectively.
Clear peaks of $\omega$ and $\phi$ are seen in the $e^+e^-$ mass spectra,
as well as peaks of $\phi$ in the $K^+K^-$ mass spectra.
The solid lines are the best fits of sums of mesons' components.
While relatively large modification is predicted on $\omega$,
there is inseparable $\rho$ contribution
in this mass region of the $e^+e^-$ invariant mass spectra.
$\phi$ is a cleaner probe,
despite relatively small modification expected,
because there is no overlapping contribution of other resonances.
In case of $d$+Au, there is a more combinatorial background.
Although with limited statistics,
peaks of $\omega$ and $\phi$ in $e^+e^-$ and $\phi$ in $K^+K^-$
are seen in the mass spectra.
In case of Au+Au, while there is a huge background,
a peak of $\phi$ is seen in the $K^+K^-$ invariant mass spectra,
and the peaks of $\phi$ and $\omega$ are seen in $e^+e^-$
after precise subtraction of the combinatorial background.
Though we may apply line-shape analysis
if we had more statistics,
it is difficult with the present data samples.

\begin{figure}[tb]
 \centering
 \begin{minipage}{0.4\hsize}
  \subfigure[]{
\includegraphics[clip,width=1.0\linewidth]{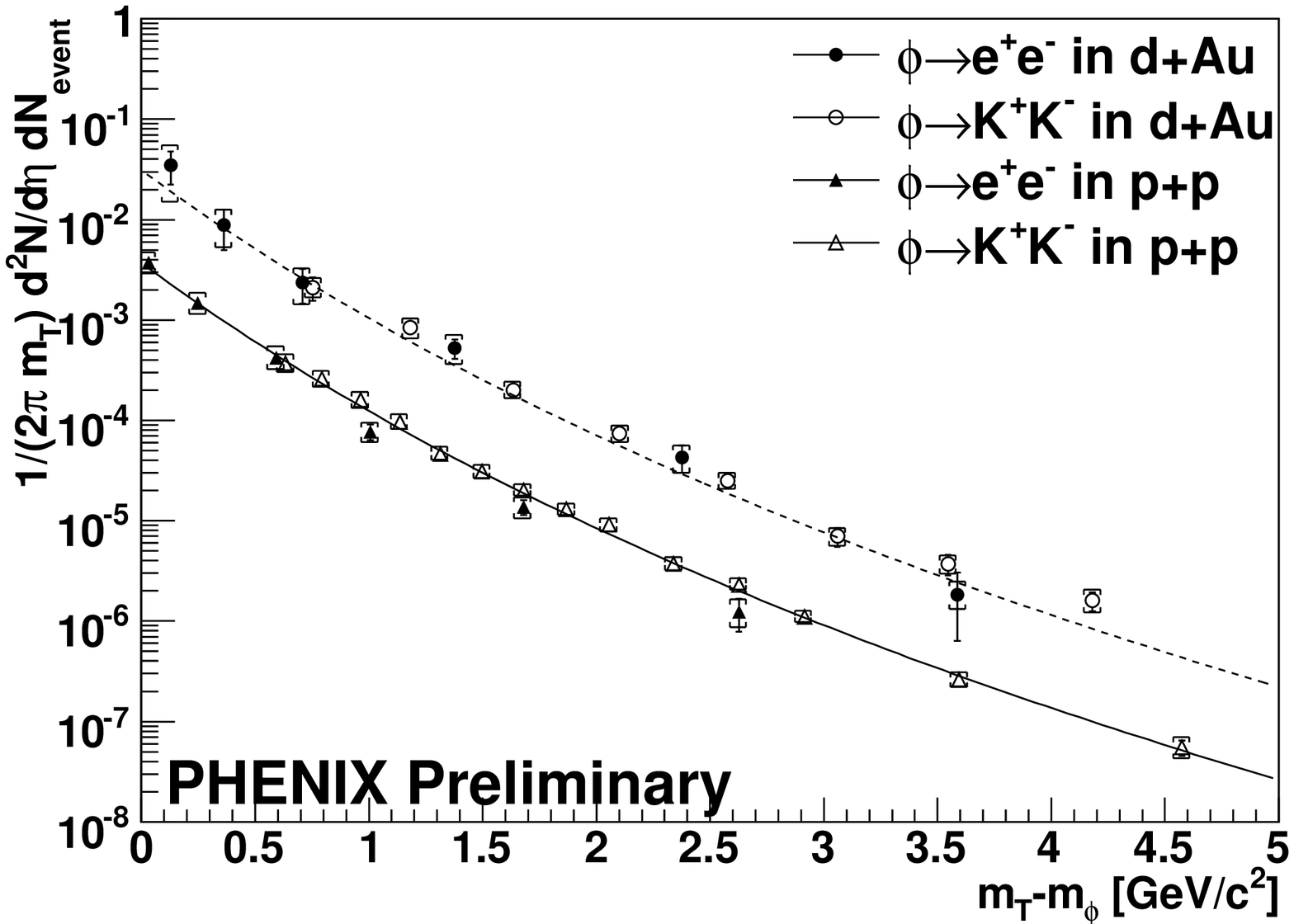}
  \label{fig:mT_pp_dAu}
  }
 \end{minipage}
 \begin{minipage}{0.1\hsize}
  \hspace{1cm}
 \end{minipage}
 \begin{minipage}{0.4\hsize}
  \subfigure[]{
\includegraphics[clip,width=1.0\linewidth]{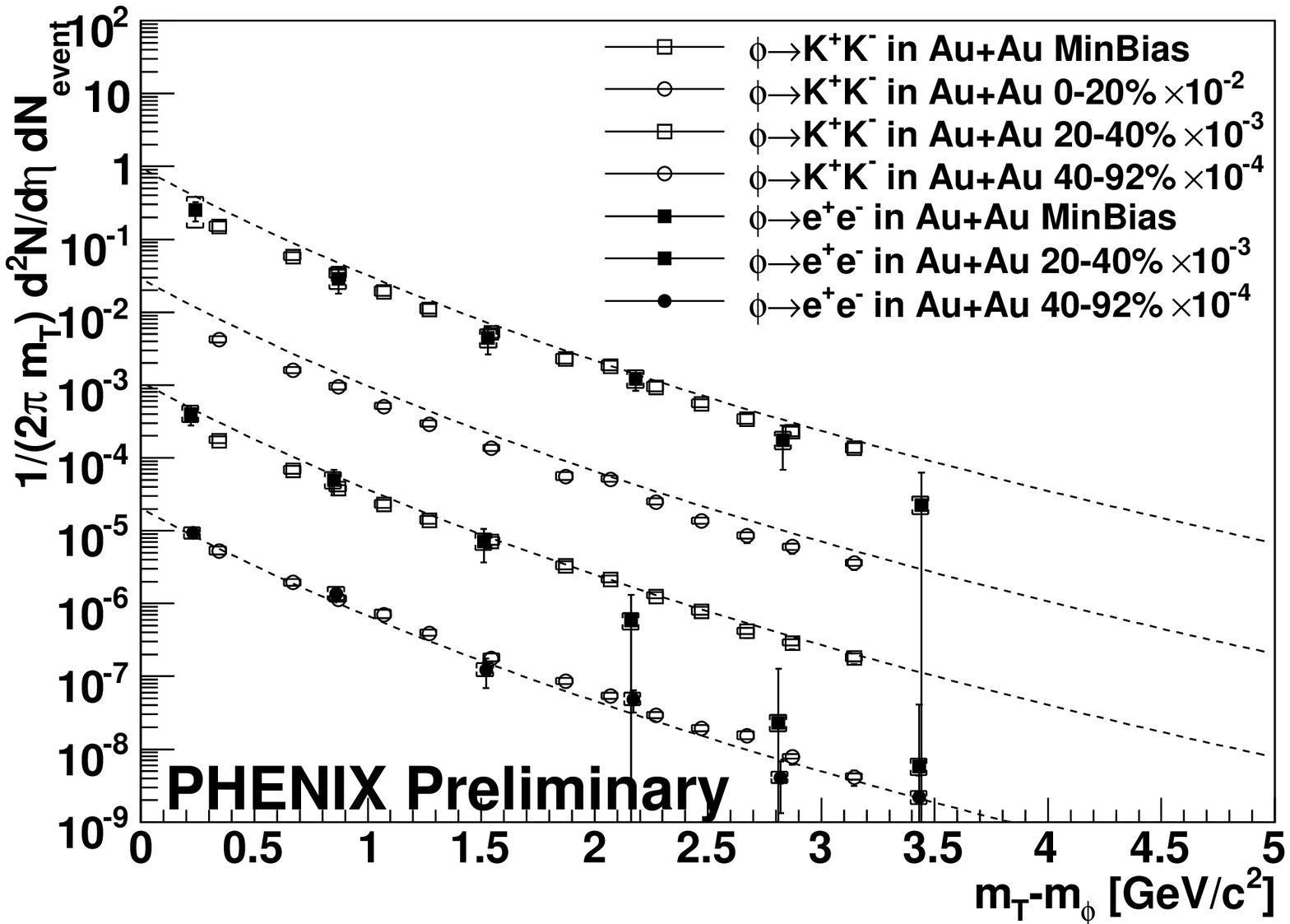}
  \label{fig:mT_AuAu}
  }
 \end{minipage}
\caption{
 Corrected $\phi\rightarrow e^+e^-$ and $\phi\rightarrow K^+K^-$
 spectra as measured in (a) $p$+$p$ and $d$+Au and (b) Au+Au collisions
 at $\sqrt{s_{\rm NN}}$ = 200~GeV.  
 The solid line in (a) is a fit to a Levy function of both channels in p+p collisions.
 The dashed lines in (a) and (b) are the fit scaled by $N_{\rm coll}$ for the case of d+Au and Au+Au.
 The $N_{\rm coll}$ scaled p+p is a reasonable fit to the d+Au spectra.
 }
 \label{fig:mt-spectra-hh+ee-pp+dAu+AuAu-200}
\end{figure}

The number of $\phi$ and $\omega$ are counted in the mass regions,
and corrected for the acceptance and efficiency.
Figure \ref{fig:mt-spectra-hh+ee-pp+dAu+AuAu-200} shows
the corrected $m_{\rm T}$ spectra of $\phi\rightarrow e^+e^-$
and $\phi\rightarrow K^+K^-$
(a) in $p$+$p$ and $d$+Au collisions
and (b) in each centrality of Au+Au collisions,
where the solid lines are global fits
for $\phi\rightarrow K^+K^-$ and $e^+e^-$,
and dashed lines show the yield in $p$+$p$
scaled by the number of binary nucleon-nucleon collisions ($N_{\rm coll}$)
for each collision centrality.
The spectra of $\phi\rightarrow e^+e^-$ and $\phi\rightarrow K^+K^-$
are consistent with each other in all collisions.

\begin{figure}[tb]
 \centering
 \begin{minipage}{0.3\hsize}
  \subfigure[]{
\includegraphics[clip,width=1.0\linewidth]{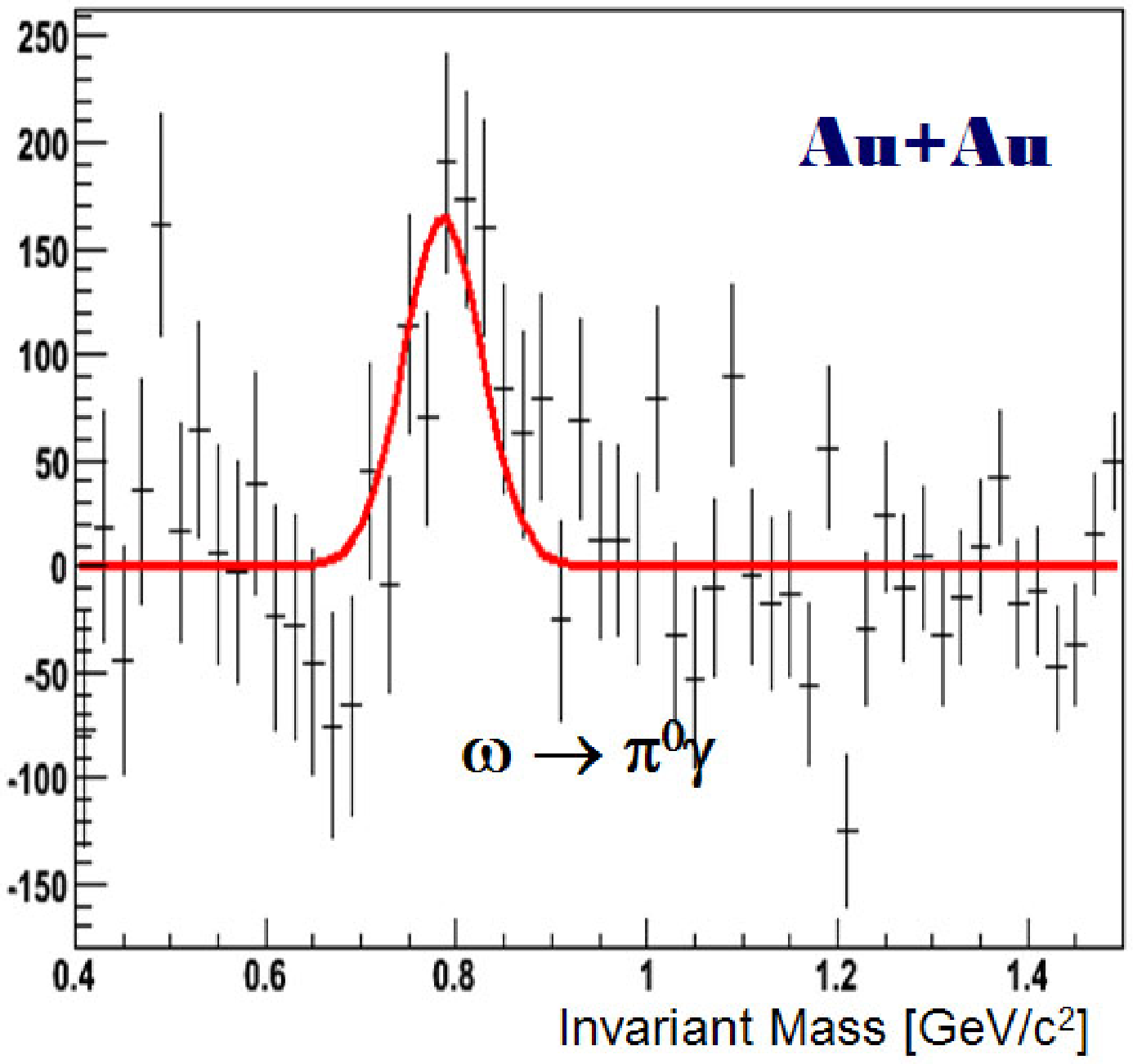}
  \label{fig:pi0gamma_AuAu}
  }
 \end{minipage}
 \begin{minipage}{0.33\hsize}
  \subfigure[]{
\includegraphics[clip,width=1.0\linewidth,height=0.9\hsize]{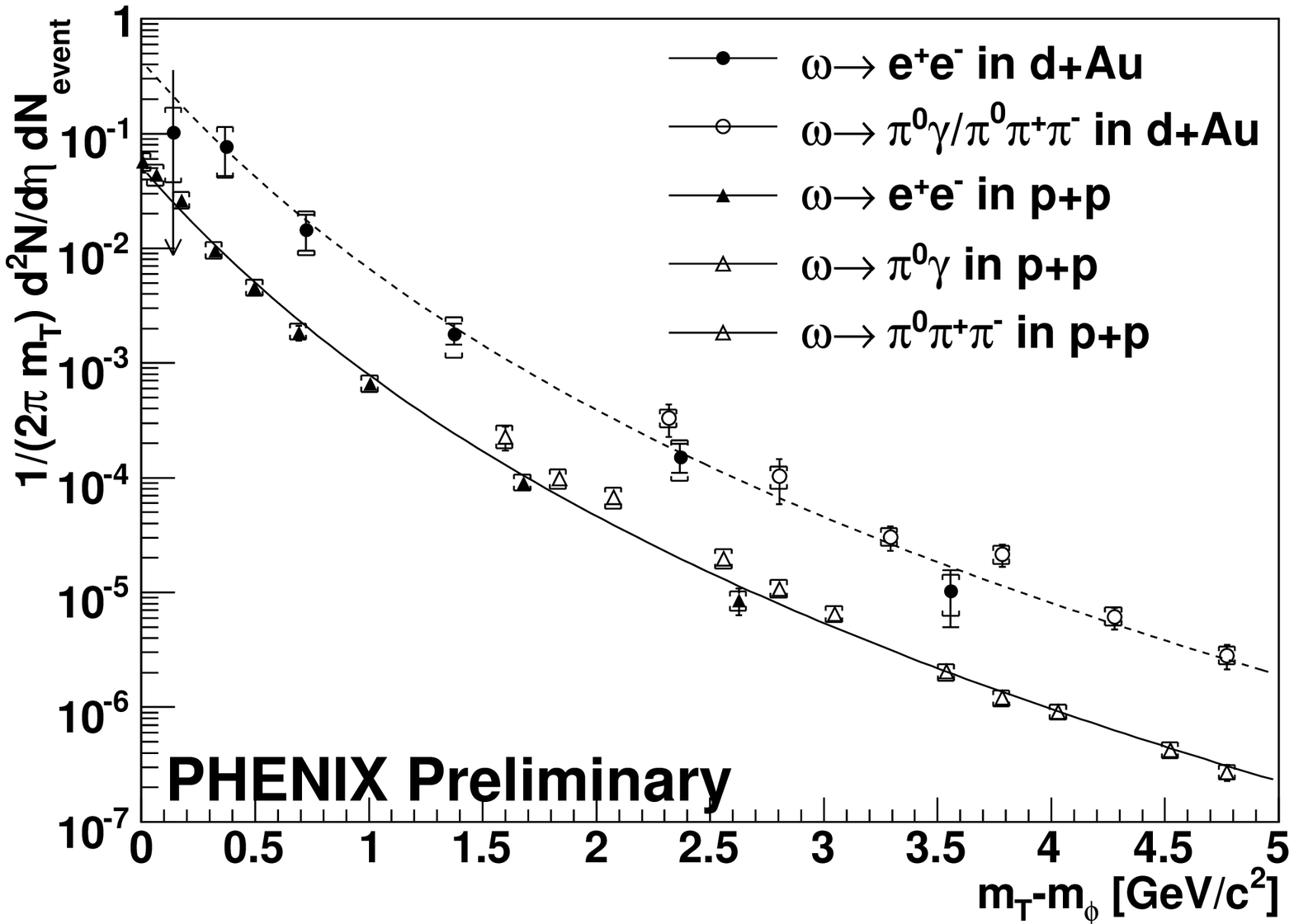}
  \label{fig:mT_omega_pp_dAu}
  }
 \end{minipage}
 \begin{minipage}{0.33\hsize}
  \subfigure[]{
\includegraphics[clip,width=1.0\linewidth]{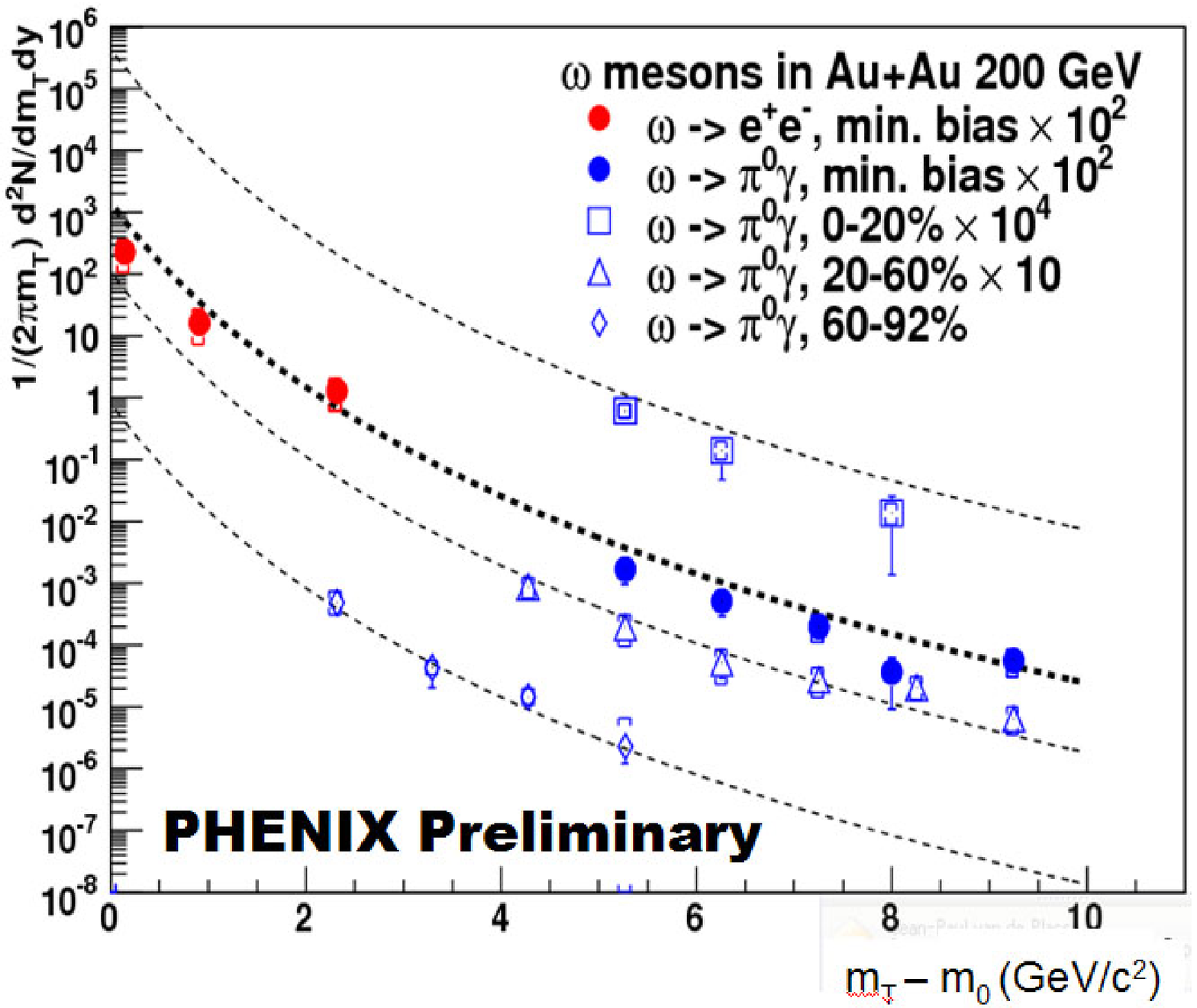}
  \label{fig:mT_omega_AuAu}
  }
 \end{minipage}
 \caption{
 Invariant mass spectra of $\pi^0\gamma$ in Au+Au collisions (a) and
 m$_{\rm T}$ spectra of $\omega\rightarrow\pi^0\gamma$ and $e^+e^-$
 in (a) $p$+$p$, $d$+Au and (c) Au+Au collisions.
 A solid line in (b) is a global fit by Levy function for the spectra in p+p collision.
 Dashed lines in (b) and (c) are scaled this fit by $N_{\rm coll}$ for d+Au and Au+Au.
 \label{fig:omega}
 }
\end{figure}

In case of $d$+Au, the yield is slightly higher than the scaling line.
This is a well-known nuclear effect called Cronin effect.
In case of Au+Au,
the yield is suppressed in the most central collisions
at least in the $K^+K^-$ channel.
The details of the $\phi$ suppression is reported
in another talk~\cite{Naglis:2009uu}.
As mentioned in the introduction,
the branching ratio of $\phi\rightarrow K^+K^-$ to $e^+e^-$
is an important measurement.
A change of the ratio may be seen as a difference in these spectra.
Something interesting may be expected
at low momentum in the central collisions.
More statistics and careful analysis are needed
to discuss a possible difference in the central collisions.

We also measured $\omega\rightarrow e^+e^-$ and $\pi^0\gamma$.
Figure \ref{fig:omega} (a) shows an invariant mass spectra
of 3$\gamma$ in Au+Au collisions.
$\omega$ is seen at high $p_{\rm T}$ in Au+Au collisions
in the $\pi^0\gamma$ channel.
Figure \ref{fig:omega} (b) and (c) show corrected $m_{\rm T}$ spectra
of $\omega\rightarrow e^+e^-$ and $\omega\rightarrow\pi^0\gamma$
in $p$+$p$, $d$+Au and Au+Au collisions,
where solid line is a global fit for $\omega$ in $p$+$p$,
and dashed lines show the yield in $p$+$p$
scaled by $N_{\rm coll}$ for each centrality.
They agree with the $N_{\rm coll}$ scaling,
while suppression may be seen in the most central collisions.

\section{Summary and Outlook}

In-medium modification of low-mass vector mesons is
one of the best probes to study chiral symmetry restoration
in hot and/or dense matter.
We measured productions of $\phi$ and $\omega$ mesons
via electronic and hadronic decay channels
in $p$+$p$, $d$+Au and Au+Au collisions at $\sqrt{s_{\rm NN}}$ = 200~GeV/$c$
in PHENIX\@.
The production is consistent with each other in the both decay channels,
and agree with the $N_{\rm coll}$ scaling.
These systematic measurement methods for the LVM's are
highly established in PHENIX\@.
A higher statistics run with a HBD detector,
which has a capability to reject the $\pi^0$ dalitz background,
will enable measurements in central Au+Au collisions
and is planned at the end of this year.
The data of the run may increase
the accuracy of the LVM measurement significantly.




\end{document}